**D.V. Vlasov, L.A. Apresian, V.I. Krystob, T.V. Vlasova**

# On the mechanisms of switching between two steady states of electroconductivity in plasticized transparent PVC films

The results of experimental researches of electroconductivity of PVC films, plasticized by patented modifier, at fields below the breakdown level are described. A possibility of management of the repeated transitions between two states with high and relatively low conductivity with conservation of reversibility is found out. A simple qualitative model of abnormal conductivity based on representation of polymer film as a sequence p-n of transitions is offered. For samples of plasticized PVC films with thickness 30-50 microns specific volume resistance of steady states was of an order of $10^4$ Om*m и$10^6$ Om*m, accordingly. A simple qualitative model describing abnormal character of conductivity of polymeric films is offered. The model considers the presence of known non-uniform plasticized polymer structure with discrete domains in which quasi-free moving of charges can occur. At imposing of an external field semi-condictive domains form sequence of p-n transitions which provides presence of two states of conductivity, in analogy with dinistors.

Приведены результаты экспериментальных исследований электропроводности поливинилхлоридных (ПВХ) пленок, пластифицированных патентованным модификатором типа «А», при напряженностях поля значительно меньших уровня пробоя. Обнаружена возможность управления переходом полимерной пленки в состояние высокой проводимости при условии сохранения многократности и обратимости переходов между двумя состояниями - с высокой и относительно низкой проводимостью: для образцов пленок пластифицированного ПВХ толщиной 30-50 мкм удельные объемные сопротивления устойчивых состояний составляли порядка $10^4$ Ом*м и $10^6$ Ом*м, соответственно. Предложена простая качественная модель, описывающая аномальный характер проводимости полимерных пленок. Модель учитывает наличие известной неоднородной структуры пластифицированного полимера с дискретными доменами, внутри которых может осуществляться квази-свободное перемещение зарядов. При наложении внешнего поля полу-проводящие домены образуют последовательность p-n переходов, которая и обеспечивает наличие двух состояний проводимости, в частности, по аналогии с динисторами.

В большинстве опубликованных исследований по переключению состояний электропроводности полимеров речь идет прежде всего о тонких пленках, причем это обстоятельство активно принимается во внимание при попытках анализа механизмов переключения состояний. [1,2]. Приводящие к изменениям электропроводности полимерных пленок явления типа эффектов аномальной близости, электронного переключения, электроформовки, «мягкого пробоя» и фриттинга специфичны именно для тонких пленок с толщиной меньше одного микрона [2]. С другой стороны во многих приложениях используются «толстые» пленки с толщиной более 10 микрон, для которых указанные эффекты мало исследованы. Более того, «толстые» полимерные пленки, в том числе пленки из пластифицированного ПВХ [3], который является наиболее широко распространенным изолятором, в настоящее время применяются повсеместно. Представляется, что изучение особенностей проводимости таких пленок важно не только с точки зрения выяснения физических механизмов возбуждения токов «утечки» [4] и переноса зарядов в полимерных пленках, но также и для многообразных практических приложений ПВХ изоляционных материалов.

В работе [5] приведены результаты экспериментальных измерений проводимости пластифицированных ПВХ пленок толщиной более 30 мкм, которые обнаруживают спонтанные переходы в состояние с высокой проводимостью (СВП), происходящие как при переключении напряжения так и при постоянном приложенном напряжении. Все измерения выполнялись при относительно низких напряжениях - почти на порядок ниже табличных значений порога пробоя (напряженность поля в полимере не превышала 2 В/мкм).

Для получения отсчетов напряжения и тока, как и в [5], в качестве образца использовалась стандартная кольцевая ячейка от ГОСТированного прибора Е6-13 с полной заменой измерительной части на автоматизированный комплекс на базе микропроцессора C8051F120 с встроенным 12 разрядным АЦП и программно управляемым предусилителем при быстродействии ядра процессора до 100 МГц. Интервал отбора отсчетов тока и напряжения составлял 1-2 мин и превосходил все характерные времена (1/RC) установления в схеме измерительного комплекса, что позволило детально отслеживать и компенсировать как релаксационные процессы, так и скачки проводимости образца под действием приложенного поля. Пленки ПВХ изготавливались как со стандартным пластификатором диоктилфталатом (ДОФ), так и с использованием нового пластификатора типа А, предложенного в [3], методом полива на стеклянные плоские подложки из 4% раствора ПВХ с пластификатором в тетрагидрофуране (ТГФ). Соотношение ПВХ и пластификатора составляло 100:80 (вес.%, соответственно). Описанные ниже исследования проводимости наблюдались в образцах ПВХ пленок, пластифицированных модификатором типа «А», и не наблюдались в образцах с ДОФ [5].

В настоящей работе показано, что приложение импульсных перепадов напряжения стимулирует переход в СВП - устойчивое состояния с удельным сопротивлением на три и более порядка ниже, чем у пленки в исходном состоянии. В отличие от описанных в литературе для тонких полимерных пленок скачкообразных изменений электропроводности при воздействии разнообразных факторов, которые упоминались выше, в наших экспериментах прикладываемые поля были на порядок ниже пороговых, а переходы происходили как на фронте роста так и при спаде напряжения. Обратный переход из СВП в исходное состояние, как правило, осуществлялся спонтанно, а наблюдаемое время жизни могло изменяться от единиц минут до нескольких часов.

Следует подчеркнуть, что импульсный перепад напряжения, подаваемый в этих экспериментах, уменьшал приложенное напряжение, и, таким образом, наблюдаемый переход в СВП не связан с предпробойными эффектами или приближением прикладываемого напряжения к порогу пробоя. Отсюда можно заключить, что наблюдаемое переключение инициируется именно перепадом напряжения, а не его величиной.

С целью анализа возможностей управляемого переключения образца в СВП была реализована специальная схема, позволявшая подавать на исследуемый образец разовый импульсный перепад напряжения – с последующим стандартным измерением отклика электро- сопротивления образца. В такой схеме удалось реализовать близкую к 100% вероятность переключения образца в СВП и отслеживать его последующий спонтанный переход в исходное состояние. Характерная зависимость изменений сопротивления образца ПВХ пленки пластифицированной модификатором «А» в логарифмическом масштабе приведена на Рис. 1.

СВП наблюдается при каждом импульсном перепаде приложенного напряжения. Моменты приложения импульсного напряжения на Рис. 1 обозначены вертикальными стрелочками. Представляется возможным найти управляющий сигнал и для обратного переключения состояния проводимости в исходное состояние, однако на настоящем этапе измерялись характерные времена жизни СВП.

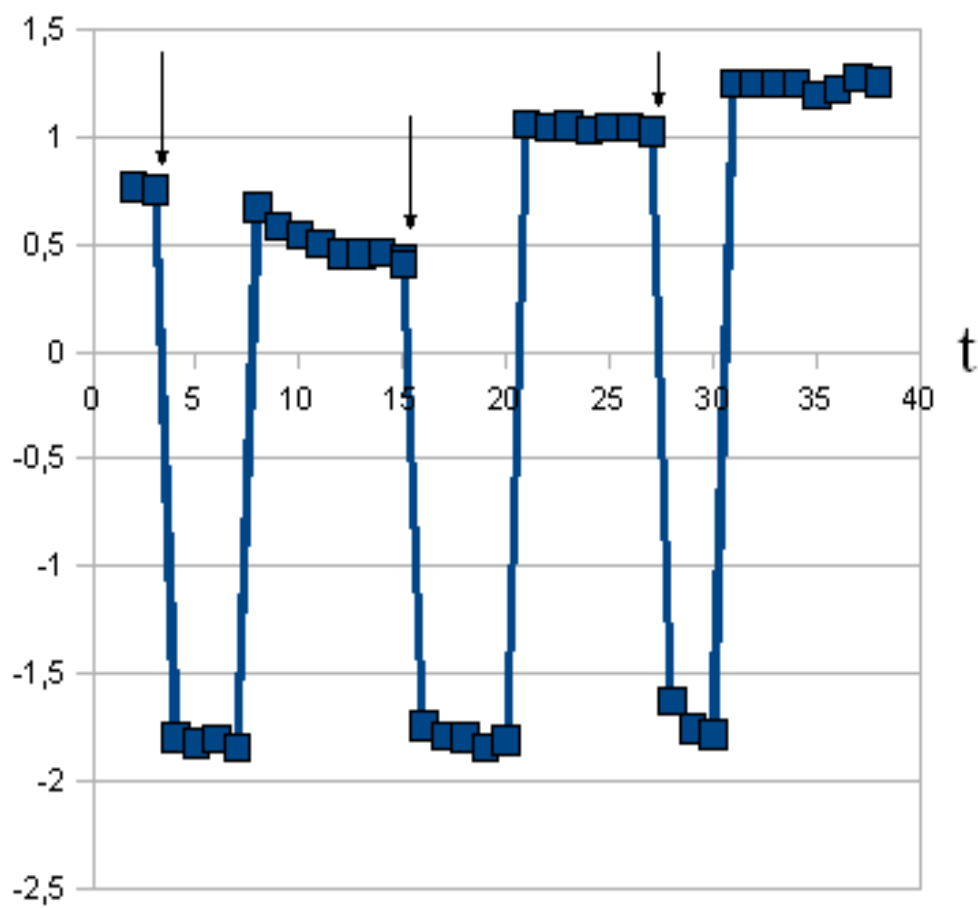

Рис. 1. Зависимость сопротивления образца (в логарифмическом масштабе) от времени. Стрелками на графике обозначены точки приложения импульсных перепадов напряжения.

Хотя строгая физическая картина наблюдаемых явлений в настоящее время отсутствует (этот факт отмечался в недавнем обзоре [2] применительно ко всем широкозонным полимерам), для объяснения перехода в СВП вспомним, что пластифицированный ПВХ имеет сложную пространственно неоднородную микро- и макро- молекулярную структуру и содержит микроскопические квази-кристаллические и аморфные образования, состоящие, в свою очередь из различных по строению молекулярных объектов (атактической, синдиотактической и изотактической природы).

Такое сложное строение позволяет говорить [2] о наличии «полупроводящих» доменов с размерами порядка размеров молекулы полимера, на стыках которых могут формироваться обсуждавшиеся в литературе каналы электронной перескоковой проводимости. Разумно предположить, что пластификатор играет роль, аналогичную роли допантов в случае ненасыщеных электропроводящих полимеров, что приводит к возникновению ограниченного числа зарядов, которые могут относительно свободно перемещаться в пространственно ограниченных областях (доменах) полимера и перескакивать посредством туннелирования в соседние домены. Длительные (до десятков минут) релаксационные процессы «установления» внутренних полей в полимере могут свидетельствовать о том, что квази- свободные заряды постепенно концентрируются на границах доменов, образуя двойные зарядовые слои или виртуальные p-n переходы. Таким образом, согласно предлагаемой гипотезе полимерную пленку во внешнем поле можно рассматривать как набор последовательных p-n переходов, причем увеличение внешнего поля приводит к более сильному «запиранию» структуры, и соответственно, уменьшению тока туннелирующих через переходы электронов, наблюдавшуюся в [5]. Такая модель - подобия полимерной пленки и приборов типа динистора – семистора, позволяет по аналогии объяснить наличие СВП и возможность многократного обратимого переключения состояний проводимости. В рамках рассматриваемой модели можно предположить, что при определенных условиях в образце реализуется аналог лавинного неразрушающего пробоя, т.е. туннелирующий через p-n переход электрон, проникая в новый домен порождает два и более свободных заряда, таким образом осуществляется переход в проводящее состояние.

Наблюдавшийся в экспериментах переход в СВП при перепадах напряжения также укладывается в рамки модели , поскольку при резком изменении внешнего поля «внутренние» напряженности поля между доменами сохраняются (времена релаксации – минуты), причем ускорение электрона в этих полях также может привести к возникновению эффекта умножения электронов и переходу в СВП даже при резком понижении прикладываемого поля.

Отметим, что экспериментальные результаты, полученные в рамках настоящей работы и предлагаемая интерпретация могут оказаться полезными для дальнейшего развития физических моделей проводимости полимерных пленок. Для пластифицированного модификатором «А» пленок ПВХ такие измерения выполнены впервые.

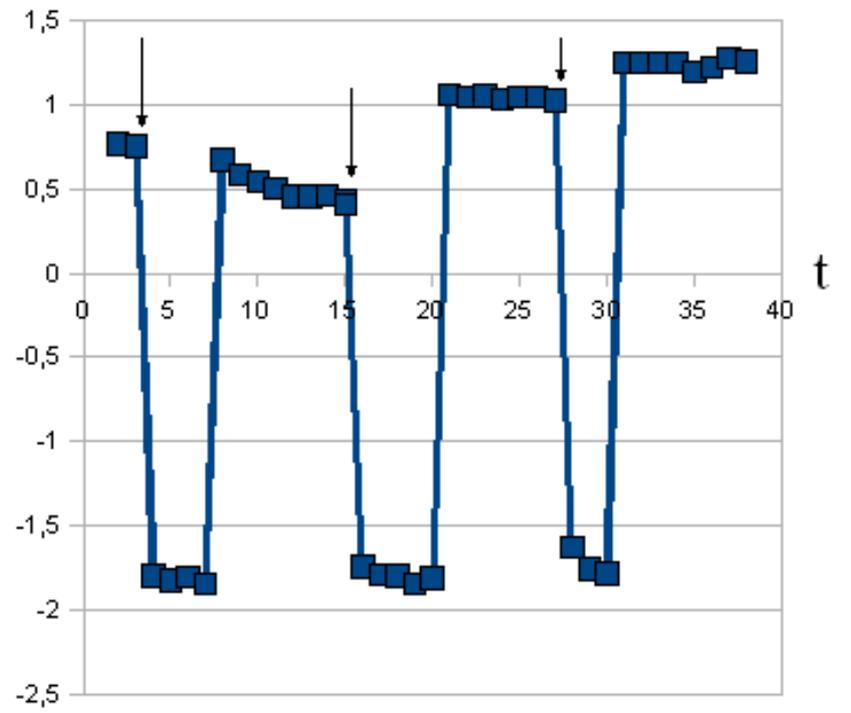